\documentclass{article}
\pdfoutput=1
\usepackage{graphicx}
\usepackage{epsfig}
\usepackage{amssymb}
\usepackage{upgreek,bm}

\title{Moments Preserving and high-resolution Semi-Lagrangian Advection Scheme \thanks{This work was partially supported by Fondo Sectorial CONACYT-SENER grant number 42536 (DGAJ-SPI-34-170412-217)}} 

\author{Juli\'an Becerra-Sagredo \thanks{(juliansagredo@gmail.com)}
\and Carlos M\'alaga \footnotemark[5] \thanks{(cmi.ciencias@ciencias.unam.mx)}
\and Francisco Mandujano\footnotemark[5] \thanks{(frmas@ciencias.unam.mx)}}

\begin{document}
\maketitle

\renewcommand{\thefootnote}{\fnsymbol{footnote}}
\footnotetext[5]{Physics Department, School of Science, Universidad Nacional Aut\'onoma de M\'exico}

\renewcommand{\thefootnote}{\arabic{footnote}}

\begin{abstract}
We present a forward semi-Lagrangian numerical method for systems of transport equations
able to advect smooth and discontinuous fields with high-order accuracy.
The numerical scheme is composed of an integration of the transport equations along the 
trajectory of material elements in a moving grid and a reconstruction of the fields in a 
reference regular mesh using a non-linear mapping and adaptive moment-preserving interpolations.
The non-linear mapping allows for the arbitrary deformation of material elements. Additionally, interpolations
can represent discontinuous fields using adaptive-order interpolation near 
jumps detected with a slope-limiter function. Due to the large number of operations during the interpolations, a 
serial implementation of this scheme is computationally expensive. The scheme has been accelerated in many-core 
parallel architectures using a thread per grid node and parallel data gathers. 
We present a series of tests that prove the scheme to be an attractive option for simulating advection 
equations in multi-dimensions with high accuracy.

\end{abstract}

\pagestyle{myheadings}
\thispagestyle{plain}
\markboth{J. Becerra-Sagredo, C. M\'alaga and F. Mandujano}{Semi-Lagrangian Advection Scheme}

\section{Introduction}

Lagrangian transport schemes refer to the use of the method of characteristics for the integration of 
partial differential equations with advective derivatives. When used as numerical schemes,  
the deformation of the discrete fields could lead to the loss of mesh connectivity and
non-desirable resolution variability. 
These problems can be fixed if the method is combined with periodic regularizations or interpolations 
to a fixed mesh, producing what is known as a semi-Lagrangian scheme \cite{Courant,Sawyer,Staniforth}. 

Semi-Lagrangian (SL) transport schemes are among the less restrictive numerical
solvers of advection-dominated
fluid flow problems, allowing much larger time steps than Eulerian-based advection.
They are designed to have the enhanced stability of Lagrangian trajectories and the regular resolution 
of an Eulerian grid.  This is accomplished by using a grid either as the initial or 
final configuration for every time step, integrating paths either forward or backward in 
time and using interpolations to reconstruct the fields when needed. 

Backward SL schemes are well known in meteorology. They have evolved slowly from 1952 to 1972   
\cite{Fjo1,Wii,Krish,Sawyer,Purnell},  
and became widely used three decades ago after the work of 
of Robert \cite{Robert1,Robert2}. His work was of great importance because the combination
of  a backward SL scheme with 
semi-implicit time stepping allowed for one order of magnitude longer time steps than previous schemes,
without loosing 
accuracy and became the basis of many operational 
models for weather prediction and environmental safety \cite{Pudykiewicz1989,Tanguay1990,Rasch1991}. 
Since then, many improvements to the method have been presented for the modeling of the atmosphere 
[see \cite{Staniforth} for a review]. 
The backward SL scheme has been formulated for different number of levels \cite{McDonald1,Temperton1},
and has been
made mass-conservative, for divergence-free flows,  
using finite-volumes \cite{Roache,Lin,Tanaka,Xiao,Nakamura}, finite-elements \cite{Morton,Cote}
and limiters \cite{Lentine2011}, or with 
special interpolating \cite{Bermejo} and non-interpolating procedures \cite{Ritchie1,Ritchie2}. 

Forward SL schemes have been previously studied in meteorology. 
Interesting results were produced by Purser and Leslie \cite{Leslie}
for a mass-conserving scheme  and the 
globally conservative second and third-order scheme using
cascade interpolation presented by Nair {\it et al.},  
\cite{Nair}. Cascade interpolation 
is an efficient serial procedure but
it is not multi-dimensional nor moment-preserving \cite{Purser1}.
Also, there are related forward-trajectory Lagrangian particle methods that represent the fluid 
as a field of particles that carry a non-deforming 
basis function with them \cite{Gingold,Monaghan}. In SL methods, nodes are not particles but the vertices 
of deformable material elements and should take into account variations of the volumes 
around convergent or divergent points in the flow.

Forward SL schemes have been used successfully in plasma physics to solve the 
Vlasov equation with phase space particle density functionals
\cite{Sonnen,Crous,Crous2}. They evolved from 
a close relative, the particle-in-cell (PIC) scheme \cite{Eastwood},
that suffered, like any other particle method, from 
noisy fields. In the PIC scheme, the electric and magnetic fields are computed in a grid, after a 
deposition or remapping of the particle strengths. The method was improved considering the 
particle field as a continuum and using interpolations to reconstruct the particle field into the grid 
after a fix number of time steps, keeping the regular resolution and controlling the noise.
The PIC and the 
SL schemes have been combined in \cite{Vadlamani}, obtaining a forward SL 
approach that allows reconstructions of the global field after a certain number of time steps,
further reducing non-desirable 
damping effects due to frequent interpolations. 

The forward SL scheme studied in this work was first presented
for smooth fields in the context 
of Large-Eddy Simulations (LES) \cite{Becerra-Sagredo}. 
It has evolved from Vortex-in-cell schemes \cite{Cottet} and the 
Method of Transport (MoT) \cite{Fey}. Its main contribution was the introduction of novel
high-order moment preserving 
explicit interpolation formulas named Z-splines which are the collocation basis of finite
differences, constructed using 
Hermite interpolation \cite{BS1}. 
This forward SL scheme was applied to reactive flows using the complete 
three-dimensional compressible Navier-Stokes equations 
with a reduced combustion model showing almost null numerical diffusion and dispersion.

The main concerns with SL schemes are the low-order iterative backward integration of trajectories, 
the lack of formal conservation properties, particularly for compressible fields, and
the use of computationally expensive high-order interpolations.
So far, high-order integration in time has been achieved using forward
integrations combined with reconstructions over successive lines 
or with multi-dimensional maps. Formal conservation has been proved only for
incompressible flows using finite-volume flux corrections, 
Galerkin schemes or high-moments conservative interpolations. While computationally
expensive interpolations have remained a 
fundamental problem until  the arrival of the fine-grain parallelism of many-core architectures like
the graphics processing units (GPUs). 

Here, we have extended the third-order in time forward SL scheme of \cite{Becerra-Sagredo}  
to allow the advection of discontinuous fields in multi-dimensions, 
automatically adjusting the interpolation basis around the discontinuity using a slope-limiter criteria. 
The result is a high-order advection scheme that transports smooth fields without losses in 
several high-order moments of its distributions while detecting jumps and kinks in the
transported fields. Interpolations represent fields near discontinuities with lower order functions to avoid 
using data across jumps, and suppress spurious oscillations.
The algorithm has been parallelized for
many-core architectures, using an independent process for 
each node of the grid. The high-order interpolations are accelerated using the
many-core wide memory bandwidth gather operations. 

In the next section a detailed description of the numerical scheme is presented. The methods 
performance and behavior under a variety of tests is shown in the third section. The last section
summarizes the work presented in this paper and comments on the applications and further development of the 
scheme we are currently working on.

\section{High-order Forward semi-Lagrangian Transport}

We consider the system of $N$ transport equations in one, two or three 
dimensions, 
using Einstein's notation, 
 given by 
\begin{equation}
\frac{D}{Dt} { h_i} ({\boldsymbol x},t) \equiv  \frac{\partial}{\partial t} { h_i} +
u_j \frac{\partial}{\partial x_j} { h_i} = S_i( {\boldsymbol h} , {\boldsymbol x} ,t),
\label{momento}
\end{equation}
where $j = 1,...,N$ and the velocity vector 
\begin{equation}
u_j = \frac{d x_j}{ dt}
\label{vel}
\end{equation}
defines the equation of motion for a material trajectory $x_j =x_j \left( {\boldsymbol x}_0,t \right)$ 
with initial position
 $x_{j,0} = x_j(t=0)$ and $ \frac{D}{Dt}$ is the material derivative.  
 
The integration in time of equations (\ref{momento}) and (\ref{vel}) is possible knowing the velocity
field as a function of time, yet being given or as part of the transported quantities $h_i$, directly
or indirectly. 

This is implemented by defining a regular reference regular grid as the initial setup of a deformable
Lagrangian grid.
The fields evolving on the Lagrangian grid are interpolated back to the reference grid after a certain 
number of time steps, avoiding highly deformed material elements and reseting the Lagrangian grid and
corresponding integration paths. High-order interpolations are needed when reseting the Lagrangian
grid to avoid the introduction of errors higher than spatial and temporal discretizations.  

The right hand side (RHS) of equation (\ref{momento}) may contain spatial derivatives. This are computed
using finite differences on the reference grid. To do so, fields must be interpolated from the
Lagrangian to the reference grid. Same interpolations are used to reset the Lagrangian grid to
the reference. Time evolution is computed using a low storage, third order Runge--Kutta
\cite{Williamson} over a number of time steps before interpolating the required fields on the
reference grid. The interpolation strategy is presented in the following section. 

\subsection{Z-spline interpolations}

The interpolation procedure chosen for the construction of a multi-dimensional function from an 
orthogonal data set, consists of tensor products of an special type of one-dimensional piecewise Hermite 
polynomials called Z-splines that are the collocation basis of finite differences. 
To the authors knowledge, this procedure was first introduced in 2003 \cite{BS1, Becerra-Sagredo}. 

The Z-spline $Z_{m}(x)$ interpolating the 
data $(x_{i},h_{i})$, $i =1,...,n$ with $x_1 < x_2 < ... < x_n$, is a  
piecewise polynomial function which satisfies the conditions:
\begin{equation}
\label{eq:zcontin}
Z_{m}(x) \in C^{m}([x_{1},x_{n}]), 
\end{equation}
\begin{equation}
\label{eq:matchder}
\left. \frac{d^{p}}{dx^{p}}Z_{m}(x) \right|_{x_{j}} = h_{m,j}^p, 
\quad \hbox{for} 
\begin{array}{l}
p = 0,...,m \\ \; j = 1,...,n
\end{array}
\end{equation}
\begin{equation}
\label{eq:zpolynom}
\qquad \qquad Z_{m}(x) \in \pi_{2m+1}([x_{i},x_{i+1}]) \quad \hbox{for} \; i = 1,...,n-1;
\end{equation}
where $h_{m,j}^p$ is the $p$th order derivative of $h$ obtained by high-order finite differences 
of $h_{j}$ using the Taylor series from $2m+1$ points as centered as possible, 
and $\pi_{n}$ stands for the class of polynomials of degree not exceeding $n$, over 
the field $\mathbb{R}$ of real numbers.

For a regular grid,  $x$ is scaled with the grid spacing $\Delta x$ and, 
without loss of generality, one can assume the grid width to be $1$.  
Cardinal splines, in general, are special cases of spline interpolations given 
the data points $x_{j} = j$ for $j \in \mathbb{Z}$. In this situation, the 
expressions for the Z-splines simplify because all cardinal basis functions have the same form and differ 
only by integer shifts. 

Let $\widetilde{Z}_{m}(x)$ be the basis function for 
cardinal Z-splines interpolating the data. With this definition the interpolation formula 
for general data in $\mathbb{R}$ can be written as
\begin{equation}
Z_{m}(x) = \sum_{i = -\infty}^{\infty} h(x_{i}) \widetilde{Z}_{m}(x-x_{i}).
\end{equation}

The linear, cubic and quintic cardinal Z-splines are given in Appendix A. 
Important properties of the cardinal Z-splines are: 

(a) $\widetilde{Z}_{m}(x) = 0$ for  $|x| > m+1$, {\it i.e.}, has compact support.

(b) The first $2m+1$ discrete moments of the 
cardinal Z-spline basis function $\widetilde{Z}_{m}$ are conserved, {\it i.e.},
for $n = 0,...,m$ we have
\begin{equation}
\sum_{j = -\infty}^{\infty} (x-j)^{n} \widetilde{Z}_{m}(x-j) = \delta_{0n} 
= \left\{ 
\begin{array}{ll}
1&   $if$ \;  n=0, \\
0&   $otherwise$.
\end{array}
\right.
\end{equation}

(c) For sufficiently smooth functions, the cardinal 
Z-spline basis functions are $L_{2}$ accurate to order $2m+1$.

Multi-dimensional Z-spline interpolations in a regular Cartesian mesh are the sum of the products of the
neighboring node functional values, with the product of the cardinal Z-splines in every Cartesian
direction. Consider a field $h(x,y,z)$ defined on a three dimensional Cartesian space and represented
by a set of discrete values $h_{i.j,k} = h(x_i,y_j,z_k)$ on a Cartesian grid. Let $(x,y,z)$ be a point
such that $x_r < x < x_{r+1}$, $y_s < y < y_{s+1}$ and $z_t < z < z_{t+1}$ where $r$, $s$ and $t$ are
integer indexes that enumerate the points of the Cartesian grid. The interpolation of order $m$ of 
$h(x,y,z)$ is given by
\begin{equation}
 Z_m (x,y,z) = \sum_{i=r-m}^{r+m+1} \sum_{j=s-m}^{s+m+1} \sum_{k=t-m}^{t+m+1} h_{i.j,k} 
 \widetilde{Z}_{m}(x-x_i) \widetilde{Z}_{m}(y-y_j)  \widetilde{Z}_{m}(z-z_k).
 \label{multiZ}
\end{equation}
In a SL scheme the nodes move and therefore is necessary to map them into a regular 
Cartesian mesh before performing the interpolations. This is discussed in the next subsection.

\subsection{Non-linear material element mapping and localization}

In this work, high-order interpolations were performed using tensor products
of one-dimensional functions in Cartesian grids. General interpolations in non-orthogonal 
Lagrangian grids are possible using a map to a local Cartesian coordinate system. In the mapped 
space, the Lagrangian grid is Cartesian in the vicinity of a point of interest in the reference grid, now 
deformed, where the values of the fields are computed through interpolations based on equation (\ref{multiZ}).

To perform interpolations at a point on the reference Cartesian grid, the position of the point in the 
mapped space is required and can be calculated through a Taylor 
expansion of the mapping function. Furthermore, if the shape of the deformed material elements, 
represented by the Lagrangian grid, are approximated by Z-splines (\ref{multiZ}), then Taylor series are
finite and nonlinear, and can be inverted to obtain relative positions in the mapped space. Higher order 
Z-spline representation of the deformed material elements has proven to achieve little 
improvements compared to using $\widetilde{Z}_{0}(x)$ in the product (\ref{multiZ}) \cite{Becerra-Sagredo}. All 
results shown in this paper where obtained using $\widetilde{Z}_{0}(x)$ to approximate the inverse of the mapping 
function.

\begin{figure}
\centering
\setlength{\unitlength}{0.1\textwidth}
  \begin{picture}(5,5)
    \put(0,0){\includegraphics*[width=6cm]{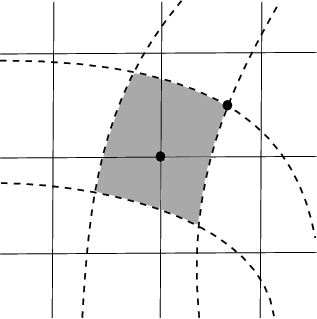}}
    \put(2.5,2.5){${\boldsymbol x}_0'$}
    \put(3.1,3.4){${\boldsymbol x}_0$}
  \end{picture}
\caption{A two-dimensional deformed Cartesian grid represented by the dashed lines. A point ${\boldsymbol x}_0'$ 
in the reference Cartesian grid, represented by the continuous lines, inside the shaded material element is 
shown. The distance to a vertex ${\boldsymbol x}_0$ of the material element, in the mapped space where the 
material element is transformed into a square, is needed for interpolating fields. }
\label{fig:element}
\end{figure}

Let ${\boldsymbol x}'$ represent the points of the Cartesian reference grid and ${\boldsymbol x}$ the points
of the Lagrangian grid. Take ${\boldsymbol x}_0'$ to be the point where interpolations are required and 
${\boldsymbol x}_0$ one of its nearest Lagrangian neighboring points. This means that ${\boldsymbol x}_0'$
is enclosed in one of the material elements of which ${\boldsymbol x}_0$ is a vertex, as shown in figure
\ref{fig:element} by the shaded deformed element.

Let ${\boldsymbol \xi} ( {\boldsymbol x} )$ be the transformation function of the deformed material
coordinate system ${\boldsymbol x}$ to a Cartesian coordinate system in the vicinity of ${\boldsymbol x}_0'$.
The inverse, ${ {\boldsymbol x} (\boldsymbol \xi} )$, can be approximated using $Z_0({\boldsymbol \xi})$
in (\ref{multiZ}) 
and the positions of ${\boldsymbol x}_0$ and the rest of vertices of the Lagrangian element containing
${\boldsymbol x}_0'$. In this way, the position of ${\boldsymbol \xi}_0' =  {\boldsymbol \xi}( {\boldsymbol x}_0' )$
inside the material element in the mapped space can be obtained implicitly through the Taylor expansion
\begin{eqnarray}
\Delta  x_i = \frac{\partial x_i}{\partial \xi_j}({\boldsymbol \xi}_0) \Delta \xi_j + \frac{1}{2} \sum_{j \neq 
k} \frac{\partial^2 x_i }{\partial \xi_j \partial \xi_k } ( {\boldsymbol \xi}_0) \Delta \xi_j \Delta \xi_k  
\label{eq:mapping1} \\  
+ \frac{1}{6} \sum_{j \neq k \neq l} \frac{\partial^3 x_i }{\partial \xi_j \partial \xi_k 
\partial \xi_l}({\boldsymbol \xi}_0)  \Delta \xi_j \Delta \xi_k \Delta \xi_l,
\nonumber
\end{eqnarray}
\noindent where $\Delta {\boldsymbol x} = {\boldsymbol x} - {\boldsymbol x}_0'$,
$\Delta {\boldsymbol \xi} = {\boldsymbol \xi} - {\boldsymbol \xi}_0'$, and the points $ {\boldsymbol x}$ are the
vertices of the material element. Finally, as derivatives in equation (\ref{eq:mapping1}) can be computed, the 
system can be inverted numerically using Newton's method to obtain $\Delta {\boldsymbol \xi}$ and perform 
interpolations. The initial condition for Newton's method is obtained by solving (\ref{eq:mapping1}) ignoring all 
the non-linear terms. 

The possibility of resetting the fields after a number of time steps require the iterative search of a 
material element. 
We initialize the search on the opposite direction of the velocity vector and, if the mapping produces 
coordinates 
out of the unitary cube, we continue the search on the neighboring cube indicated  
by the coordinates of the resulted out-of-bounds mapping, repeating the search until the bounded mapping 
and therefore
the  material element is found.  

\subsection{Discontinuities}
\label{discont}

\begin{figure}[t]
\centering 
\includegraphics[width=8cm]{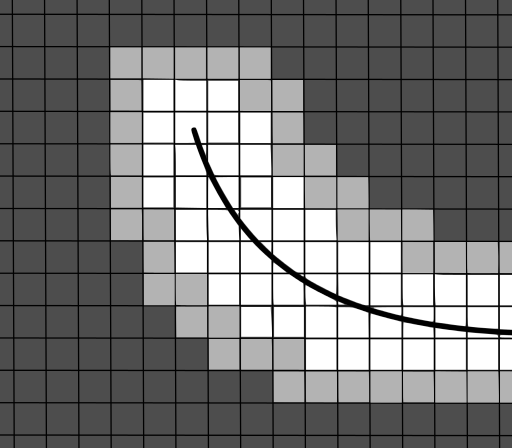} 
\caption{Diagram of the order of the interpolation in the presence of a discontinuity, represented by 
the curved line, on a Cartesian grid.
The dark gray shaded area represents the region where $Z_2$ quintic interpolation is used. Light gray is 
where  $Z_1$ cubic interpolation is used. Inside white squares, linear interpolation is used.}
\label{limiter}
\end{figure}

In order to work with discontinuous distributions and shocks appearing in advection equations, a limiter 
like jump criterion was introduced. Lets consider a one dimensional discrete distribution $h_i= h(x_i)$ for the
sake of simplicity, extension to many dimensions is straight forward. At each time step, a material point 
$x_i$ is marked with a flag whenever the following criterion is met:
\begin{equation}
\frac{| h_{i+1} -2 h_i + h_{i-1} | }{max \left( |h_{i+1} - h_i | ,|h_i - h_{i-1} | \right) } > 0.25.
\end{equation}
This represents a comparison of curvatures and slopes that provide a measure of the steepness of
the distribution at the material point. In this way, marked points are taken as the approximated 
location of a the discontinuity. 
 
When interpolations are required, flags are searched among all 
points needed for the interpolation. If a flag is found, the order of the interpolation is reduced and 
so the
number of points in the vicinity needed for interpolation, to exclude flagged points. The procedure 
is repeated until linear interpolation is reached or
no flagged points can be found in the vicinity required by the interpolation. In this way, as the nodes 
approach the discontinuity, the interpolation will only use information in one side of the discontinuity. 
This 
is represented in figure \ref{limiter} for the case of the numerical scheme with  $Z_2$ interpolation. 
Linear interpolations are used in elements lying at the jump location while cubic interpolations are used 
when the jump lies a node away.
 
This procedure naturally introduces numerical dissipation at nodes close to the discontinuities. 
As the number of nodes flagged near the discontinuity is increasingly localized when spatial resolution is 
improved, dissipation is considerably reduced with mesh refinement. 
 
\section{Results}

\begin{figure}[t]
\centering 
\includegraphics[width=12cm]{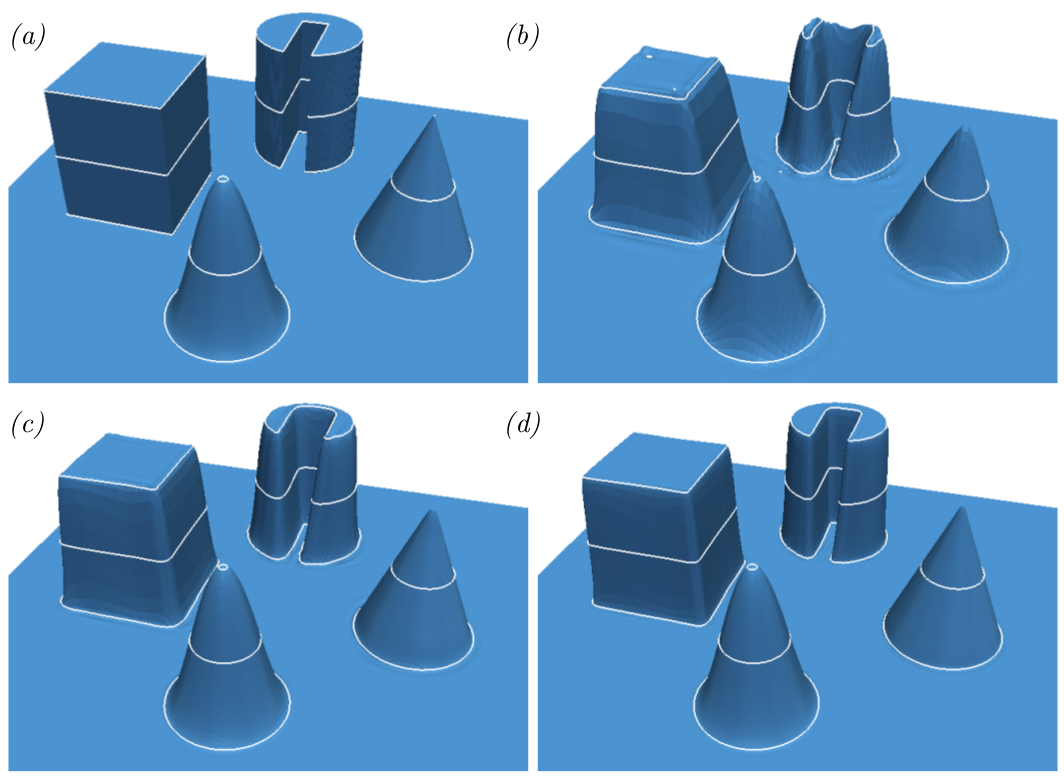} 
\caption{Solid body rotation test after 1 turn. The initial condition is plotted in (a). Solutions after one 
turn are shown in (b) for a mesh of $128 \times 128$ grid points, in (c) for $256 \times 256$ and in (d) for 
 $512 \times 512$. Contours at values of $h=0.01,0.5,0.99$ are shown.}
\label{unav}
\end{figure}
 
In this section we present a series of tests that show the method's performance. First, we solve the pure 
advection of passive scalar distributions in two dimensions. A cosine hill, a conical and 
square distributions, and a Zalesak disc are advected by the velocity field of a solid body rotation. 
Solid body rotation tests are not often shown and reveal numerical diffusion and dispersion produced 
when computing material derivatives \cite{Kuzmin-Turek2004, Kuzmin2009, Lentine2011}. Pure advection
also serves to test the moment preserving properties of the interpolation. Burger's equation in two
dimensions is computed to test advection with shock tracking.

Non-linear 
shallow water equations are used to compute the dispersion relation of linear waves, the error 
convergence in the case of Kelvin waves when a Coriolis term is included, and the method's 
convergence for nonlinear Kelvin wave dispersion. Kelvin waves are traveling wave solutions for the Coriolis 
term, in the vicinity of the equator, in the linear limit (small amplitudes). They behave as passive scalar 
fields in pure advection for small amplitudes, but start shedding equatorial Poincar\'e waves as amplitude 
increases. Therefore, they provide a test of the scheme behavior in both limits.

\subsection{Pure advection}

The advection of a passive, two dimensional scalar distribution $h(x_1,x_2)$ by a solid body rotation 
around the center of the domain $[0,1] \times [0,1]$ given by the velocity field ${\boldsymbol u} = 
\Omega( 0.5  -  x_2, x_1- 0.5)$ is a solution of the equation
\begin{equation}
 \frac{D}{Dt} h = \frac{\partial}{\partial t} h + {\boldsymbol u} \cdot \nabla h = 0,
\end{equation}

\noindent where $\Omega$ represents the angular velocity of the rotation. The initial condition, shown 
in figure \ref{unav}(a), represents a square and a conical distributions along a cosine hill and a 
Zalesak disc (the notched cylinder), all of unitary height. Figures \ref{unav}(b), (c) and (d) 
correspond to surface plots of the solutions for $\Omega=1$ after one turn, for meshes of $128 \times 
128$, $256 \times 256$ and $512 \times 512$ grid points respectively.

Contours of values $h=0.01$, $0.5$ and $0.99$ are shown on the surface plots of figure \ref{unav}. No 
noticeable distortion of the continuous cosine hill and conical distribution is found after one turn, 
showing the method computes correctly the advective derivatives of continuous functions for all the 
spatial resolutions presented.

\begin{figure}
\centering
\includegraphics[width=12cm]{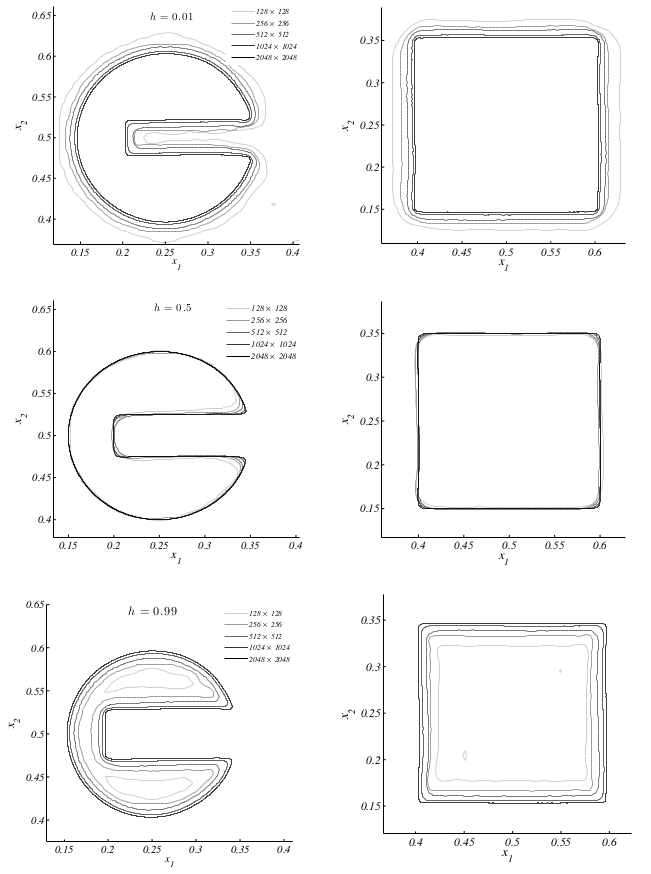}
 \caption{Contour plots for values $h=0.01,0.5,0.99$ for the discontinuous square distribution (right column)
 and 
 Zalesak disc (left column) after one turn for meshes of $128 \times 128$, $256 \times 256$, $512 \times 512$ and 
 $1024 \times 1024$ grid points.}
 \label{contornos}
\end{figure}

Contour plots of the solutions for the discontinuous square distribution and Zalesak disc after one turn 
are in figure \ref{contornos}. Contour values correspond to those of the surface plots 
($h=0.01,0.5,0.99$). Interpolation procedures adapt when grid points are in the vicinity of the 
discontinuities, reducing the order of the interpolation to avoid the use of values across the jumps in 
their computations. This induces numerical dissipation near the discontinuities, which can bee seen as 
smoothing of the distributions in figure \ref{unav}. This is a local effect that depends on the spatial 
resolution and improves considerably with it, as shown in figure \ref{contornos}. Results show that high 
resolution is needed when discontinuities appear in the process, but this represent a minor setback 
because of the efficiency of the method when implemented in massively parallel architectures as the 
GPU's. 

\begin{figure}
\centering
%
%
%
\includegraphics[width=12cm]{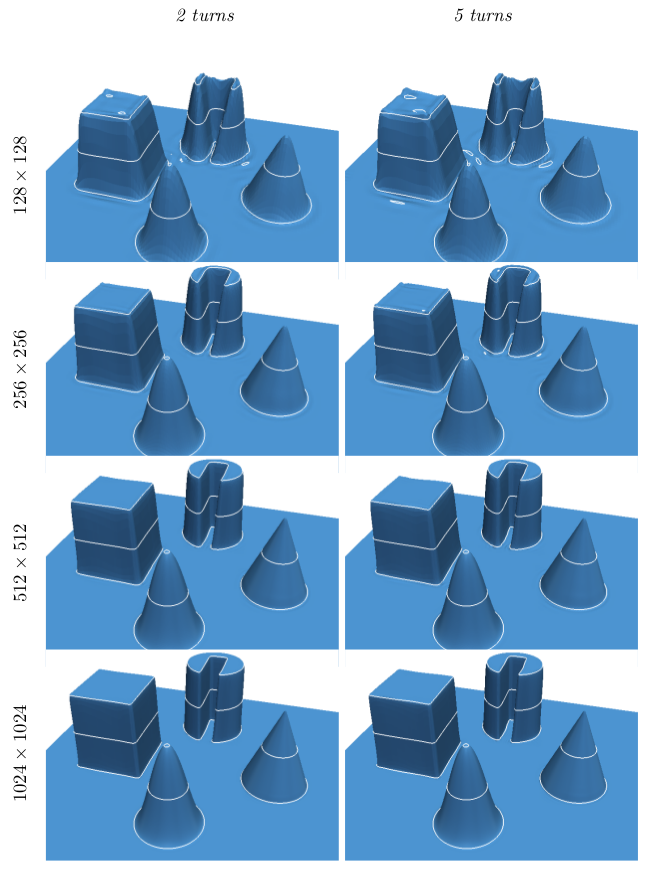}
\caption{Solid body rotation test after 2 and 5 turns. Subfigures on the left column correspond to solutions
after 2 turns, and on the right column are solutions after five turns. Plots corresponding to different number of 
grids point are organized in rows from 
$128 \times 128$ on top to $1024 \times 1024$ at the bottom.}
\label{dosycinco}
\end{figure}

Once the discontinuities are smoothed by the adaptive interpolations, distributions become continuous 
for 
a given spatial resolution and the jump criterion ceases to find a discontinuity. This means that no 
more 
numerical dissipation is produced as high order interpolations are used everywhere after smoothing. This 
can bee seen in figure \ref{dosycinco} where solutions after two and five turns are shown for different 
resolutions. Notice that continuous solutions, including the smoothed initially discontinuous, are 
advected without apparent distortion after five turns.

\begin{figure}
\centering
\includegraphics[width=10cm]{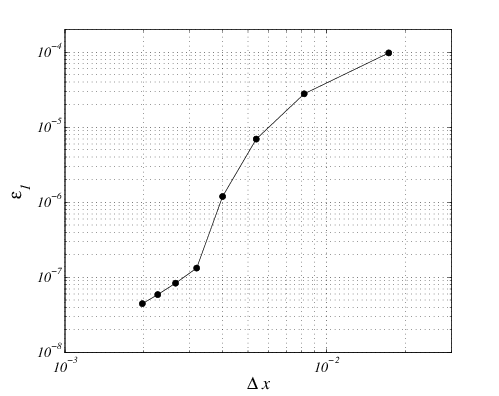} 
\caption{Relative $L_1$-error ($\epsilon_1$) after one turn for the pure advection of a cosine hill under a solid 
body rotation velocity field. }
\label{Error_SBR}
\end{figure}

The relative $L_1$ error for the continuous cosine hill distribution, given by the initial 
condition
\begin{equation}
\label{eq:alpha}
h(x_1,x_2) = 
\left\{
\begin{array}{@{\hspace{-0.00025in}}l@{\hspace{0.1in}}l@{\hspace{0.1in}}l}
0.5 \left[1+\cos \left(\frac{\pi \sqrt{\beta}}{0.1}
\right) \right] & \hbox{for} & 
\beta \leq 0.01, \\
0 &  &  \hbox{otherwise},
\end{array}
\right.
\end{equation}
where $\beta = (x_1 - 0.25)^{2}+ (x_2 - 0.5)^{2}$, after one turn are shown in figure 
\ref{Error_SBR}
for different spatial resolutions. The relative $L_1$-error ($\epsilon_1$) decays with a slope of $3$ roughly, 
see figure \ref{Error_SBR}. 

The dissipation error is defined as $[ \sigma ( h ) - \sigma (  h^*) ]^2 - [ \bar{h} - \bar{h^*}]^2$ where 
$\sigma$  is the standard deviation function, $h$ the initial condition \ref{eq:alpha}, $h^*$ the numeric 
solution after one turn, and the over-bar represents the mean value. For a $25 \times 25$ 
mesh, the dissipative error obtained is of approximately $10^{-5}$ with $Z_2$ interpolations, two orders of 
magnitude smaller than what is found with $Z_1$ and suitable for all the tests shown in the next section.  

Moment preserving properties of the $Z$-spline interpolations are shown in Table \ref{momentos} for the
discrete moments  defined as 
\begin{equation}
m_{p} = \sum_{i} h_{i}^{*} \left[ (x_{1}^i- 0.5)^{p} + (x_{2}^i-0.5 )^{p} \right], 
\end{equation} 
\noindent where the sum is over all points in the reference grid enumerated by $i$ and $ h_{i}^{*} $ is the 
numerical solution after one turn on the grid point $(x_1^i, x_2^i)$. The table shows $m_p$ for $p= 0,1,2,3,4$ 
of the cosine hill distribution
using the $Z_0$, $Z_1$ and $Z_2$ interpolations for a  $25 \times 25$ mesh, and their exact values. 

\begin{table}
\footnotesize
\centering
\begin{tabular}[h]{|@{\hspace{0.3in}}c|c|c|c|c|c@{\hspace{0.3in}}|}
\hline
     & $m_0$ & $m_1$ & $m_2$ & $m_3$ & $m_4$   \\
\hline \hline
Exact  &       23.3536 & -11.6768 &  6.0558 & -3.0820 & 1.6249 \\
\hline
FSL-$Z_{0}$ & 22.8278 & -11.0854 &  8.0065 & -4.1267 & 3.1567 \\
\hline
FSL-$Z_{1}$ & 23.3553 & -11.6787 &  6.0583 & -3.0991 & 1.6201 \\
\hline
FSL-$Z_{2}$ & 23.3535 & -11.6768 &  6.0559 & -3.0822 & 1.6251 \\
\hline
\end{tabular}
\caption{Discrete moments of the cosine hill used in the solid body rotation test
after one period.}
\label{momentos}
\end{table}

Finally, we comment on the scheme performance when implemented to run in GPU's. The execution
of a time step showed a linear growth from $0.02$ seconds for a $150$ thousand grid points to $0.4$ seconds 
for a $4$ million grid points on a Nvidia\textregistered TeslaK20 Card (CFL number of  0.5). This is 
characteristic of explicit schemes
under fine grain parallelism, when the time evolution of fields on a single node 
represents a computational thread.

\subsection{Burger's equation}

As seen in the previous section, passive discontinuous distributions can be advected by the FSL method 
using the jump criterion described in Section \S\ref{discont} which produced a localized numerical diffusion 
and a smoothing of the discontinuity as a consequence. Although this gives good results in pure 
advection, it is not desirable to smooth out discontinuities when solutions should produce shocks. 
Burger's equations provides a test to the jump criterion behavior under shock formation. Burger's 
equation

\begin{equation}
\frac{D}{Dt} {\boldsymbol u} = \frac{\partial}{\partial t} {\boldsymbol u} + 
{\boldsymbol u} \cdot \nabla {\boldsymbol u} = {\boldsymbol 0},
\end{equation}

\noindent is solved in the domain $[0,1] \times [0,1]$ for the initial condition

\begin{equation}
{\boldsymbol u} = 
\left\{
\begin{array}{@{\hspace{-0.00025in}}l@{\hspace{0.1in}}l@{\hspace{0.1in}}l}
 \frac{1}{r}e^{-150(r-0.2)^2} (x_1-0.5, x_2-0.5) & \hbox{for} & 
r > 0.001, \\
{\boldsymbol 0} &  &  \hbox{otherwise},
\end{array}
\right.
\end{equation}
\noindent where $r= \sqrt{(x_1-0.5)^2 + (x_2-0.5)^2}$.

\begin{figure}[ht]
\centering
\includegraphics[width=11cm]{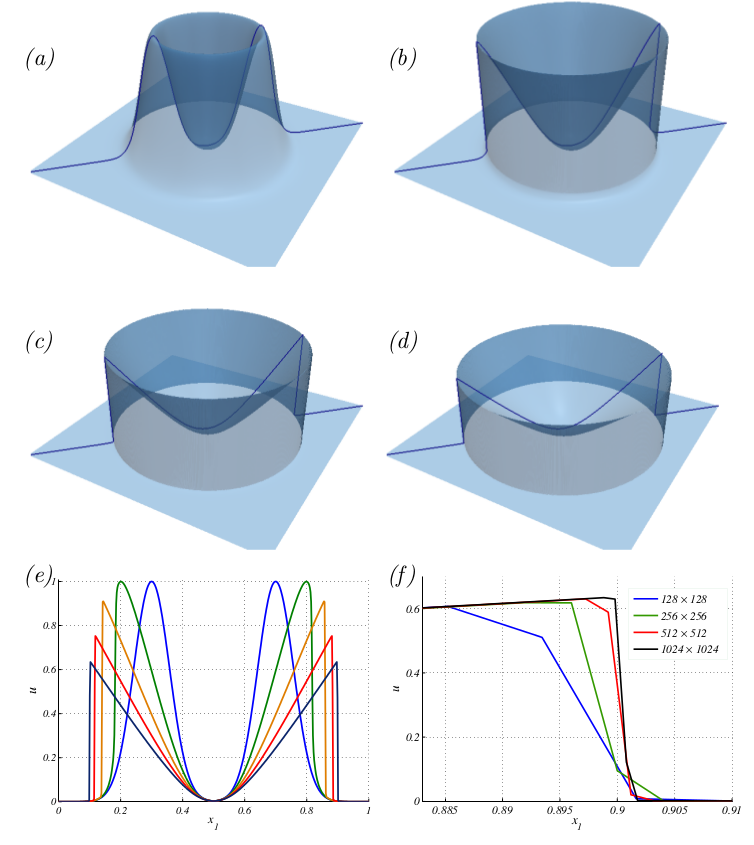}
\caption{Numerical solution of Burger's equation in two dimensions. The initial condition is shown in (a). 
Solutions at times $0.1$, $0.2$ and $0.3$ are shown in (b), (c) and (d) respectively. In (e), a radial slice
of solutions at $t=0, 0.1, 0.2, 0.3, 0,4$ are compared. A close view of the shock at $t=0.4$ for different grid 
resolutions can be seen in (f).}
\label{burgers}
\end{figure}

Shock formation can be seen in figure \ref{burgers}. Solutions show that the jump 
criterion preserves the shock without apparent diffusion or oscillations in the vicinity of the front. Figure 
 \ref{burgers}(f) suggest convergence to the actual discontinuity as resolution increases.

Results show that the jump criterion can capture and advect shocks  which makes the scheme suitable for a wider 
variety of applications.

\subsection{Non-linear shallow-water equations}

The non-linear shallow water equations (NLSW) describe the evolution of the perturbation $h$ of the 
thickness of a layer of fluid under gravity in the limit were the characteristic wavelength of the 
perturbation is much larger than the layers thickness. It is the result of vertically averaging the mass 
and momentum equations of the flow across the layer \cite{majda}. The set of equations for $h$ and the horizontal 
averaged velocity ${\boldsymbol u}$, the advecting field, can include viscosity effects, topography 
and Coriolis forces to approximate the behavior of large structures in oceanic and atmospheric flows 
\cite{majda}.

NLSW were chosen to test the method in different ways. First, to reproduce the linear dispersion relation 
of freely oscillating gravity waves, with no viscosity or Coriolis effects. Simulation of Kelvin 
equatorial waves in the linear regime served to compute the order of the method when the Coriolis term is 
included in the equations. Convergence of the method for non-linear phenomena was tested computing 
equatorial Kelvin non-linear wave dispersion.  

The non-dimensional shallow-water equations in two spatial dimensions including Coriolis effects and 
viscosity are given by
\begin{eqnarray}
\frac{D}{Dt} h &=& - (1+h) \nabla \cdot {\boldsymbol u}, \nonumber\\
\frac{D}{Dt} {\boldsymbol u}&=& - \nabla h -f {\boldsymbol u}^\bot + \nu \nabla \cdot \left[ (1+h) (\nabla  {\boldsymbol 
u}+\nabla  {\boldsymbol u}^T) \right], 
\label{nlsw}
\end{eqnarray}
\noindent where ${\boldsymbol u}^\bot = (-u_2,u_1)$ is the orthogonal velocity \cite{majda}, $f$ is the Coriolis 
parameter proportional to $x_2$ close to the equator, placed in $x_2=0$, and $\nu$ is the 
dimensionless viscosity, taken as a small parameter.  

\begin{figure}[ht]
\centering
 \includegraphics[width=8cm]{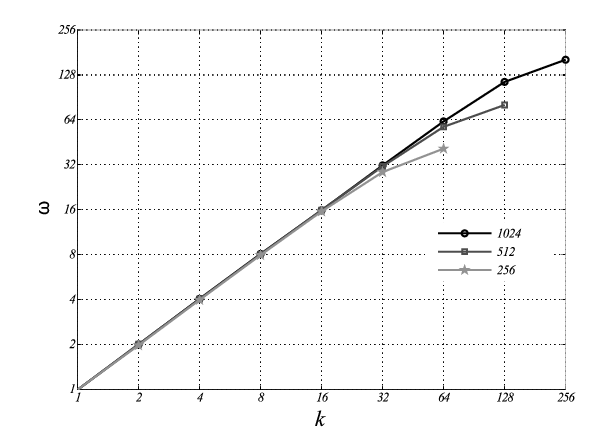} 
\caption{Linear dispersion relation for shallow water waves.}
\label{lineal}
\end{figure}

Equations were solved with bi-periodic boundary conditions over a given squared domain 
$[0,L_x]\times[-0.5,0.5]$. For the linear dispersion relations, solutions to (\ref{nlsw}) where computed for 
$f=0$, $\nu = 0$ and one spatial dimension for a small amplitude sinusoidal perturbation $h$. 
The plot of the resulting frequency $\omega$ vs the wave number $\kappa$ can be seen in 
figure \ref{lineal} for different spatial resolutions. Notice that the computed dispersion relation is 
correct until a value of $\kappa$ beyond which the spatial discretization fails to represent correctly 
the sinusoidal wave form.  

\begin{figure}[ht]
\centering
 \includegraphics[width=8cm]{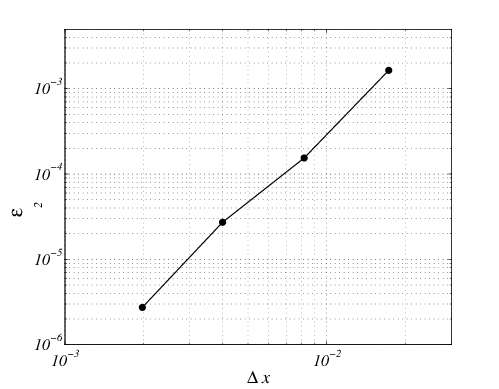}
\caption{Relative $L_2$-error ($\epsilon_2$) decay with grid refinement for the equatorial Kelvin wave simulation.}
\label{order}
\end{figure}

Equatorial NLSW equations without viscosity effects ($\nu = 0$, $f=f_0 x_2$) in the small amplitude 
(linear) limit accept an infinite set of dispersive wave solutions, called Rossby, Yanai and Poincar\'e 
waves, and non-dispersive eastward traveling wave solutions called Kelvin waves given by 
\begin{eqnarray}
h &=& e^{-f_0 x_2^2/2} K(x_1 -t),\nonumber  \\
u_1 &=& e^{-f_0 x_2^2/2} K(x_1 -t), \nonumber \\
u_2 &=& 0.
\label{kelvin}
\end{eqnarray}
\noindent Where $K$ is an arbitrary function of small amplitude \cite{majda}. To compute the order of 
the 
method we solved NLSW equations (\ref{nlsw}) for $\nu=0$, $f_0 = 250$ with an initial condition given 
by 
(\ref{kelvin}) with $t=0$ and $K(x_1) = A  e^{-f_0 x_1^2/2}$ with $A= 0.0001$. Solutions at 
dimensionless 
time $t=1$ were compared with the analytical Kelvin wave solution, the relative $L_2$-error ($\epsilon_2$) 
for different space discretization is shown in figure \ref{order}. Again, the method shows a third order 
convergence rate.

\begin{figure}[ht]
\centering
\setlength{\unitlength}{0.1\textwidth}
  \begin{picture}(7,9)
    \put(0.5,5){ \includegraphics[width=8cm]{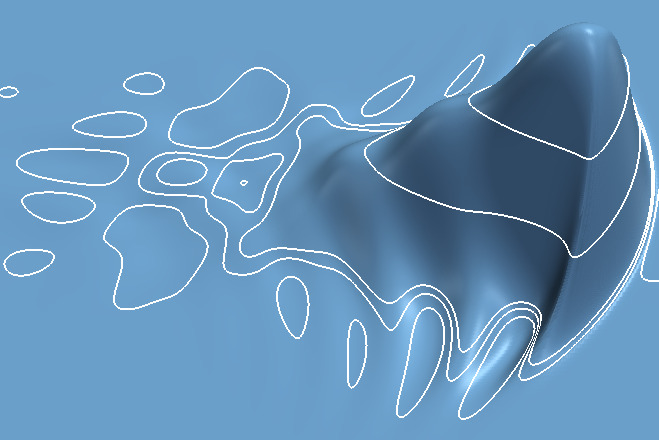} }
    \put(0.5,0){ \includegraphics[width=8cm]{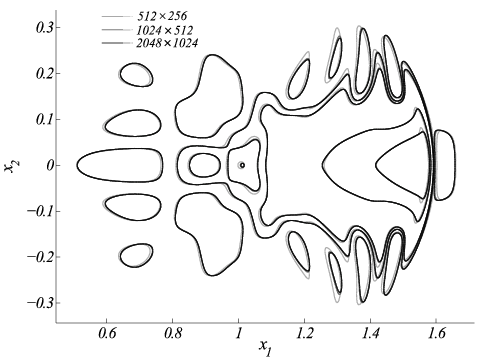} }
    \put(0,8.8){\it (a)}
    \put(0,4.2){\it (b)}
  \end{picture}  
\caption{Non-linear Kelvin-Poincar\'e wave dispersion. The surface plot of $h$ at time $t=1$ for an initial 
Gaussian bell is shown in (a). In (b) contours for the values of $h=-0.001,0.005,0.02,0.1,0.2$, corresponding to 
those in (a), are plotted for different number of grid points to show convergence. Contours of small scale 
reveal Poincar\'e wave structures on the wake.}
\label{trilobite}
\end{figure}

When computing solutions in the nonlinear regime, with initial conditions of larger amplitudes ($A 
\approx \mathcal{O} (0.1)$), perturbations produce a nonlinear excitation of dispersive Poincar\'e modes 
and the formation of a steep front; a known feature of this set of equations \cite{majda}. In figure 
\ref{trilobite} one can see the surface plot of $h$ at time $t=1$ for $\nu=0.0005$, $f_0 = 250$ and the 
same Gaussian bell initial condition for $h$ and ${\boldsymbol u}$ previously used, but with a larger amplitude 
$A=0.5$. Contours drawn on top of the surface $h$ correspond to those plotted for different resolutions 
showing convergence of the method in the nonlinear regime. Contour values given by 
$h=-0.001,0.005,0.02,0.1$ and $0.2$ were chosen to reveal the appearance of Poincar\'e modes 
characterized by oscillations in both $x_1$ and $x_2$ directions, which can be noticed in figure 
\ref{trilobite}. Also, contours show convergence of structures of all amplitudes. It most be noticed 
that a small viscosity is needed 
to show convergence as nonlinear dispersion of the initial condition will excite infinite Poincar\'e 
modes in the inviscid case, from which only a finite set can be represented by a given spatial resolution. This 
is characteristic of inviscid flow instabilities, and a dissipative mechanism must be included to introduce 
a minimum scale where flow structures can be represented by the chosen spatial resolution. 

\section{Conclusions}

In most fluid phenomena, advection plays an important roll. A numerical scheme capable of making 
quantitative predictions and simulations must compute correctly the advection terms appearing in the 
equations governing fluid flow. In many cases, as those where there are viscous effects, numerical 
errors produced by the computation of material derivatives are diffused away by dissipative mechanisms. 
Here we present a high order forward semi-Lagrangian numerical scheme specifically tailored to correctly 
compute material derivatives, as seen by its good performance under advection tests. The success of our 
scheme relies on the geometrical interpretation of material derivatives to compute the time evolution of 
fields on grids that deform with the material fluid domain, an interpolating procedure of arbitrary 
order that preserves the moments of the interpolated distributions, and a nonlinear mapping strategy to 
perform interpolations between undeformed and deformed grids. Additionally, a discontinuity criterion 
was implemented to deal with discontinuous fields and shocks. Tests of pure advection, shock formation 
and nonlinear phenomena are presented to show performance and convergence of the scheme.

On the oder hand, the high computational cost that made impractical this scheme in its origin can now be 
considerably reduced by massively parallel architectures found in graphic cards. Explicit numerical 
schemes, like the one presented here, benefit the most from such architectures, they can reduce two to 
three orders of magnitude the computation times. If we add the fact that graphic processing units give 
the best operation per monetary cost ratio, our proposed scheme is a very attractive alternative for 
advective flow simulation.

Extension to three dimensional flows is on its way and the ultimate goal is the development of a code 
for the full 3D Navier-Stokes equations for both compressible and incompressible phenomena. 

\section*{Appendix}
The $C^{1}$, cubic cardinal Z-spline basis function (equivalent to cubic Bessel interpolation) 
\begin{equation}
\widetilde{Z}_{1} (x) = \left\{
\begin{array}{lll}
1-\frac{5}{2}x^{2}+ \frac{3}{2}|x|^{3} &   $ $ & |x|\leq 1, \\
\frac{1}{2} \left( 2 - |x| \right)^{2} \left( 1-|x| \right) &
$ $ & 1 \leq |x| \leq 2, \\
0&   $ $ & |x| > 2.
\end{array}
\right.
\label{eq:z1}
\end{equation}

The $C^{2}$, quintic cardinal Z-spline basis function 
\begin{equation}
\widetilde{Z}_{2} (x) = \left\{
\begin{array}{lll}
1 - \frac{15}{12}x^{2} - \frac{35}{12}|x|^{3} + \frac{63}{12}x^{4}
- \frac{25}{12}|x|^{5} &   $ $ & |x|\leq 1, \\ [0.2cm]
-4 + \frac{75}{4}|x| - \frac{245}{8}x^{2} \\ [0.1cm] \qquad + \frac{545}{24}|x|^{3}
- \frac{63}{8}x^{4} + \frac{25}{24}|x|^{5} &  $ $  & 1\leq |x| \leq2, \\ [0.2cm]
18 - \frac{153}{4}|x| + \frac{255}{8}x^{2} \\ [0.1cm] \qquad - \frac{313}{24}|x|^{3}+ \frac{21}{8}x^{4} - \frac{5}{24}|x|^{5} &  $ $ & 2\leq |x| \leq3, \\ [0.2cm]
0 & $ $ & |x| > 3.
\end{array}
\right.
\end{equation}

{\bf Acknowledgments.} J. B. S. thanks the support of Professor Rolf Jeltsch. Authors thank Professor Antonmaria Minzoni for his comments and fruitful discussions.

\bibliographystyle{plain}

\bibliography{references}

\begin{thebibliography}{10}

\bibitem{BS1}
J.~T. Becerra-Sagredo.
\newblock {Z-splines: Moment conserving spline interpolation of compact support
  for arbitrarily spaced data}.
\newblock Technical Report 2003-10, ETH Z{\"u}rich C. H., 2003.

\bibitem{Becerra-Sagredo}
J.~T. Becerra-Sagredo.
\newblock {\em {High-order semi-Lagrangian numerical method for large-eddy
  simulations of reacting flows}}.
\newblock PhD thesis, ETH Z{\"u}rich C. H., 2007.

\bibitem{Bermejo}
R.~Bermejo.
\newblock {On the equivalence of semi-Lagrangian and particle-in-cell
  finite-element methods}.
\newblock {\em Mon. Wea. Rev.}, 118(4):979--987, 1990.

\bibitem{Cote}
J.~C{\^o}t{\'e}.
\newblock {An accurate and efficient finite-element global model for the
  shallow-water primitive equations}.
\newblock {\em Mon. Wea. Rev.}, 118(12):2707--2717, 1990.

\bibitem{Cottet}
G.~H. Cottet and P.~D. Koumoutsakos.
\newblock {\em {Vortex methods: theory and practice}}.
\newblock Cambridge University Press. U. K., 2000.

\bibitem{Courant}
R.~Courant, E.~Isaacson, and M.~Rees.
\newblock {On the solution of nonlinear hyperbolic differential equations by
  finite differences}.
\newblock {\em Comm. Pure Appl. Math.}, 52:243--255, 1952.

\bibitem{Crous2}
N.~Crouseilles, M.~Mehrenberger, and E.~Sonnendr{\"u}cker.
\newblock {Conservative semi-Lagrangian schemes for Vlasov equations}.
\newblock {\em J. Comp. Phys.}, 229(6):1927--1953, 2010.

\bibitem{Crous}
N.~Crouseilles, T.~Respaud, and E.~Sonnendr{\"u}cker.
\newblock {A forward semi-Lagrangian method for the numerical solution of the
  Vlasov equation}.
\newblock {\em Comp. Phys. Comm.}, 180(10):1730--1745, 2009.

\bibitem{Eastwood}
J.~W. Eastwood.
\newblock {The stability and accuracy of EPIC algorithms}.
\newblock {\em Comput. Phys. Comm.}, 44(1-2):73--82, 1987.

\bibitem{Fey}
M.~Fey.
\newblock {\em {Ein echt mehrdimensionales Verfahren zur L{\"o}sung der
  Eulergleichungen}}.
\newblock PhD thesis, ETH Z{\"u}rich C. H., 1993.

\bibitem{Fjo1}
R.~Fj{\o}rtoft.
\newblock {On a numerical method of integrating the barotropic vorticity
  equations}.
\newblock {\em Tellus}, 4(3):179--194, 1952.

\bibitem{Gingold}
R.~A. Gingold and J.~J. Monaghan.
\newblock {Smoothed particle hydrodynamics: theory and applications to
  non-spherical stars}.
\newblock {\em Mon. Not. R. Astron. Soc.}, 181:375--389, 1977.

\bibitem{Krish}
T.~N. Krishnamurti.
\newblock {Numerical integration of primitive equations by a quasi-Lagrangian
  advective scheme}.
\newblock {\em J. Appl. Meteor.}, 1(4):508--521, 1962.

\bibitem{Kuzmin2009}
D.~Kuzmin.
\newblock {Explicit and implicit FEM-FCT algorithms with flux linearization}.
\newblock {\em J. Comp. Phys.}, 228(7):2517--2534, 2009.

\bibitem{Kuzmin-Turek2004}
D.~Kuzmin and S.~Turek.
\newblock {High-resolution FEM-TVD schemes based on a fully multidimensional
  flux limiter}.
\newblock {\em J. Comp. Phys.}, 198(1-2):131--158, 2004.

\bibitem{Lentine2011}
M.~Lentine, J.~T. Gretarsson, and R.~Fedkiw.
\newblock {An unconditionally stable fully conservative semi-Lagrangian
  method}.
\newblock {\em J. Comp. Phys.}, 230(8):2857--2879, 2011.

\bibitem{Leslie}
L.~M. Leslie and R.~J. Purser.
\newblock {Three-dimensional mass-conserving semi-Lagrangian schemes employing
  forward trajectories}.
\newblock {\em Mon. Wea. Rev.}, 123:2551--2566, 1995.

\bibitem{Lin}
S.~J. Lin and R.~B. Rood.
\newblock {Multidimensional flux-form semi-Lagrangian transport schemes}.
\newblock {\em Mon. Wea. Rev.}, 124(9):2046--2070, 1996.

\bibitem{majda}
A.~J. Majda, R.~R. Rosales, E.~G. Tabak, and C.~V. Turner.
\newblock {Interaction of Large-Scale Equatorial Waves and Dispersion of Kelvin
  Waves through Topographic Resonances}.
\newblock {\em J. Atmos. Sci.}, 56(24):4118--4133, 1999.

\bibitem{McDonald1}
A.~McDonald.
\newblock {A semi-Lagrangian and semi-implicit two-time-level integration
  scheme}.
\newblock {\em Mon. Wea. Rev.}, 114(5):824--830, 1986.

\bibitem{Monaghan}
J.~J. Monaghan.
\newblock {Smoothed particle hydrodynamics}.
\newblock {\em Ann. Rev. Astron. Astrophys.}, 30:543--574, 1992.

\bibitem{Morton}
K.~W. Morton.
\newblock {Generalised Galerkin methods for hyperbolic problems}.
\newblock {\em Comp. Meth. Appl. Mech. Eng.}, 52:847--871, 1985.

\bibitem{Nair}
R.~D. Nair, J.~S. Scroggs, and F.~H.~M. Semazzi.
\newblock {A forward-trajectory global semi-Lagrangian transport scheme}.
\newblock {\em J. Comput. Phys.}, 190:275--294, 2003.

\bibitem{Nakamura}
T.~Nakamura, R.~Tanaka, and K.~Takizawa.
\newblock {Exactly conservative semi-Lagrangian scheme for multi-dimensional
  hyperbolic equations with directional splitting technique}.
\newblock {\em J. Comp. Phys.}, 174(11):171--207, 2001.

\bibitem{Pudykiewicz1989}
J.~Pudykiewicz.
\newblock {Simulation of the Chernobyl dispersion with a 3D hemispheric tracer
  model}.
\newblock {\em Tellus}, 41B:391--412, 1989.

\bibitem{Purnell}
D.~K. Purnell.
\newblock {Solution of the advective equation by upstream interpolation with a
  cubic spline}.
\newblock {\em Mon. Wea. Rev.}, 104:42--48, 1976.

\bibitem{Purser1}
R.~J. Purser and L.~M. Leslie.
\newblock {An efficient interpolation procedure for high-order
  three-dimensional semi-Lagrangian models}.
\newblock {\em Mon. Wea. Rev.}, 119(10):2492--2498, 1991.

\bibitem{Rasch1991}
P.~Rasch and D.~Williamson.
\newblock {The sensitivity of a general-circulation model climate to the
  moisture transport}.
\newblock {\em J. Geophys. Res.}, 96(D7):13,123--13,137, 1991.

\bibitem{Ritchie1}
H.~Ritchie.
\newblock {Eliminating the interpolation associated with the semi-Lagrangian
  scheme}.
\newblock {\em Mon. Wea. Rev.}, 114:135--146, 1986.

\bibitem{Ritchie2}
H.~Ritchie.
\newblock {Application of the semi-Lagrangian method to a spectral model of the
  shallow-water equations}.
\newblock {\em Mon. Wea. Rev.}, 116:1587--1598, 1988.

\bibitem{Roache}
P.~J. Roache.
\newblock {A flux-based modified method of characteristics}.
\newblock {\em Int. J. Num. Meth. Fluids}, 15(11):1259--1275, 1992.

\bibitem{Robert1}
A.~Robert.
\newblock {A stable numerical integration scheme for the primitive
  meteorological equations}.
\newblock {\em Atmos. Ocean.}, 19:35--46, 1981.

\bibitem{Robert2}
A.~Robert.
\newblock {A semi-Lagrangian semi-implicit numerical integration scheme for the
  primitive meteorological equations}.
\newblock {\em J. Meteorolog. Soc. Japan.}, 60:319--325, 1982.

\bibitem{Sawyer}
J.~S. Sawyer.
\newblock {A semi-Lagrangian method of solving the vorticity advection
  equation}.
\newblock {\em Tellus}, 15:336--342, 1963.

\bibitem{Sonnen}
E.~Sonnendr{\"u}cker, J.~Roche, P.~Bertrand, and A.~Ghizzo.
\newblock {The semi-Lagrangian method for the numerical resolution of the
  Vlasov equation}.
\newblock {\em J. Comput. Phys.}, 149:201--220, 1999.

\bibitem{Staniforth}
A.~Staniforth and J.~C{\^o}t{\'e}.
\newblock {Semi-Lagrangian integration schemes for atmospheric models - a
  review}.
\newblock {\em Mon. Wea. Rev.}, 119:2206--2223, 1991.

\bibitem{Tanaka}
R.~Tanaka, T.~Nakamura, and T.~Yabe.
\newblock {Constructing exactly conservative scheme in a non-conservative
  form}.
\newblock {\em Comp. Phys. Comm.}, 126(3):232--243, 2000.

\bibitem{Tanguay1990}
M.~Tanguay, A.~Robert, and R.~Laprise.
\newblock {A semi-implicit semi-Lagrangian fully compressible regional forecast
  model}.
\newblock {\em Mon. Wea. Rev.}, 118:1970--1980, 1990.

\bibitem{Temperton1}
C.~Temperton and A.~Staniforth.
\newblock {An efficient two-time-level semi-Lagrangian semi-implicit
  integration scheme}.
\newblock {\em Quart. J. Roy. Meteor. Soc.}, 113:1025--1039, 1987.

\bibitem{Vadlamani}
S.~Vadlamani, S.~E. Parker, Y.~Chen, and C.~Kim.
\newblock {The particle-continuum method: an algorithmic unification of
  particle in cell and continuum methods}.
\newblock {\em Comput. Phys. Comm.}, 164:209--213, 2004.

\bibitem{Wii}
A.~Wiin-Nielsen.
\newblock {On the application of trajectory methods in numerical forecasting}.
\newblock {\em Tellus}, 11:180--196, 1959.

\bibitem{Williamson}
J.~H. Williamson.
\newblock {Low-storage Runge-Kutta schemes}.
\newblock {\em J. Comp. Phys.}, 35:48--56, 1980.

\bibitem{Xiao}
F.~Xiao and T.~Yabe.
\newblock {Completely conservative and oscillationless semi-Lagrangian schemes
  for advection transportation}.
\newblock {\em J. Comp. Phys.}, 170(2):498--522, 2001.

\end{thebibliography}

\end{document}